# A proposition of 3D inertial tolerancing to consider the statistical combination of the location and orientation deviations

Pierre-Antoine Adragna*, Serge Samper and Maurice Pillet

SYMME Laboratory
Université de Savoie
Polytech'Savoie, BP 80439
74944 Annecy Le Vieux Cedex, France
E-mail: pierre-antoine.adragna@univ-savoie.fr
E-mail: serge.samper@univ-savoie.fr
E-mail: maurice.pillet@univ-savoie.fr
*Corresponding author

Abstract: Tolerancing of assembly mechanisms is a major interest in the product life cycle. One can distinguish several models with growing complexity, from 1-dimensional (1D) to 3-dimensional (3D) (including form deviations), and two main tolerancing assumptions, the worst case and the statistical hypothesis. This paper presents an approach to 3D statistical tolerancing using a new acceptance criterion. Our approach is based on the 1D inertial acceptance criterion that is extended to 3D and form acceptance. The modal characterisation is used to describe the form deviation of a geometry as the combination of elementary deviations (location, orientation and form). The proposed 3D statistical tolerancing is applied on a simple mechanism with lever arm. It is also compared to the traditional worst-case tolerancing using a tolerance zone.

Keywords: 3D tolerancing; inertial tolerancing; statistical tolerancing; form tolerancing; modal characterisation; statistical acceptance criterion.

Biographical notes: Pierre-Antoine Adragna is a PhD student in Mechanical Science from the Université de Savoie in France. His thesis topic is about the tolerancing of mechanisms and the fusion of the 1D inertial tolerancing and the form modal parameterisation.

Serge Samper has been a Master of Conference at the Université de Savoie since 1994. He received a PhD from the Université de Toulouse, France in 1994. His research topic is about geometrical tolerancing.

Maurice Pillet has been a Research Director at the Université de Savoie in France since 1999. He graduated from the E.N.S Cachan, France, in 1982 and received his PhD in Quality from the Université de Savoie in 1993. His research areas are statistics, experiment planning and quality.

## 1 Introduction

Tolerancing of assembly systems is a large domain. Several tolerancing approaches are proposed with varying modelling complexity. The simplest modelling corresponds to the one-dimensional (1D) tolerancing, where only dimensional variations are considered as presented by Graves (2001). Although this is the simplest tolerancing model, teams are still working on statistical methods with heterogenic and nonindependent components (Anselmetti and Radouani, 2003). Pillet (2003) proposes a statistical approach with a new criterion: the inertial tolerancing that will be used in this paper.

As this modelling may not be sufficient for complex mechanisms, three-dimensional (3D) modelling is proposed. As a nonexhaustive list of some approaches, one can distinguish the vector loop chain proposed by Chase (1999), T-maps by Davidson and Shah (2002), and the Small Displacement Torsor (SDT) used by Giordano et al. (1992), defined by Bourdet et al. (1995). This last 3D approach will be treated in this paper to express the tolerance domain associated with a tolerance zone. The tolerance values are calculated with a worst-case approach (Giordano et al., 2001), and statistical analysis (Germain and Giordano, 2007) is used.

Dealing with precision in the tolerancing of assembly systems, more complex models are proposed considering the form deviations of parts. Davidson's team proposes a model in the T-maps space to study the influence of form deviation (Ameta et al., 2007). Radouani and Anselmetti (2003) present an experimental study of the positioning regarding form deviation. Neville et al. (2006) proposes an algorithm to determine the positioning deviation of a butting assembly regarding the form deviation and roughness. Samper's team proposes an approach to determine the positioning error due to the form deviations of shapes for a given positioning force and auxiliary positioning surfaces (Adragna et al., 2007a).

The aim of this paper is to present an original 3D statistical tolerancing based on the fusion of the modal characterisation of form deviation and the inertial acceptance criterion of the 1D inertial tolerancing. This paper is a first step that only considers location and orientation deviations for the assembly of components. The article is decomposed into two main parts: a theoretical part that presents the fusion of the inertial and the modal approaches, and a part on the application of the 3D inertial tolerancing by the use of the small displacement torsor without form deviations. The assembly mechanism is a stack-up of three components with a lever arm so that the functional requirement is off-centre from the contact surfaces of the parts.

## 2 Methods used and background

This part presents three approaches:

1 inertial tolerancing and the associated inertial acceptance criterion

2 the modal parameterisation of any shapes to describe form deviations

3 the fusion of both approaches into the 3D inertial criterion.

## 2.1 The 1D inertial tolerancing

This part introduces the inertial acceptance criterion and a 1D statistical tolerance allocation called inertial tolerancing.

### 2.1.1 The inertial acceptance criterion

Inertial tolerancing concerns 1D models of mechanisms. Based on the Taguchi loss function, Pillet (2003) defines the quadratic off-centring of a batch to its target, called the batch inertia:

$$I = \sqrt{\sigma^2 + \left(\bar{X} - T\right)^2} = \sqrt{\sigma^2 + \delta^2} \quad (1)$$

where $\delta$ corresponds to the batch off-centring to its target and $\sigma$ is its standard deviation. To compare the deviation of a batch to its inertial tolerance, Pillet defines the Cpi capability index:

$$Cpi = \frac{I_0}{I} = \frac{I_0}{\sqrt{\delta^2 + \sigma^2}}. \quad (2)$$

Some compare this new criterion to the Taguchi capability index Cpm (Chan et al., 1988). The main advantage of the inertial acceptance criterion is the absence of use of a tolerance interval that usually corresponds to the acceptance limits of the parts dimension. There is then no ambiguity of acceptance of a batch included in the tolerance interval but not acceptable due to the Cpm capability index.

The representation of inertial tolerance uses two relations:

1 a tolerance on the off-centring considering the batch standard deviation:

$$\delta_{Max} = \sqrt{\left(\frac{I_0}{Cpi}\right)^2 - \sigma^2} \quad (3)$$

2 a tolerance on the standard deviation considering the mean batch off-centring:

$$\sigma_{Max} = \sqrt{\left(\frac{I_0}{Cpi}\right)^2 - \delta^2}. \quad (4)$$

Figure 1 shows a theoretical batch. The off-centring is $\delta$ = 0.010 mm, the standard deviation is $\sigma$ = 0.020 mm and the inertial tolerance is I = 0.035 mm with a capability index Cpi = 1.

Both diagrams in Figure 1 characterise the same batch. The figure on the right is the initial representation of the inertial tolerance. Because the figure on the left is closer to the traditional representation of a batch in its tolerance interval, it can be more easily understood.

Figure 1 The inertial tolerance and its,( ) graphical representation (see online version for colours)

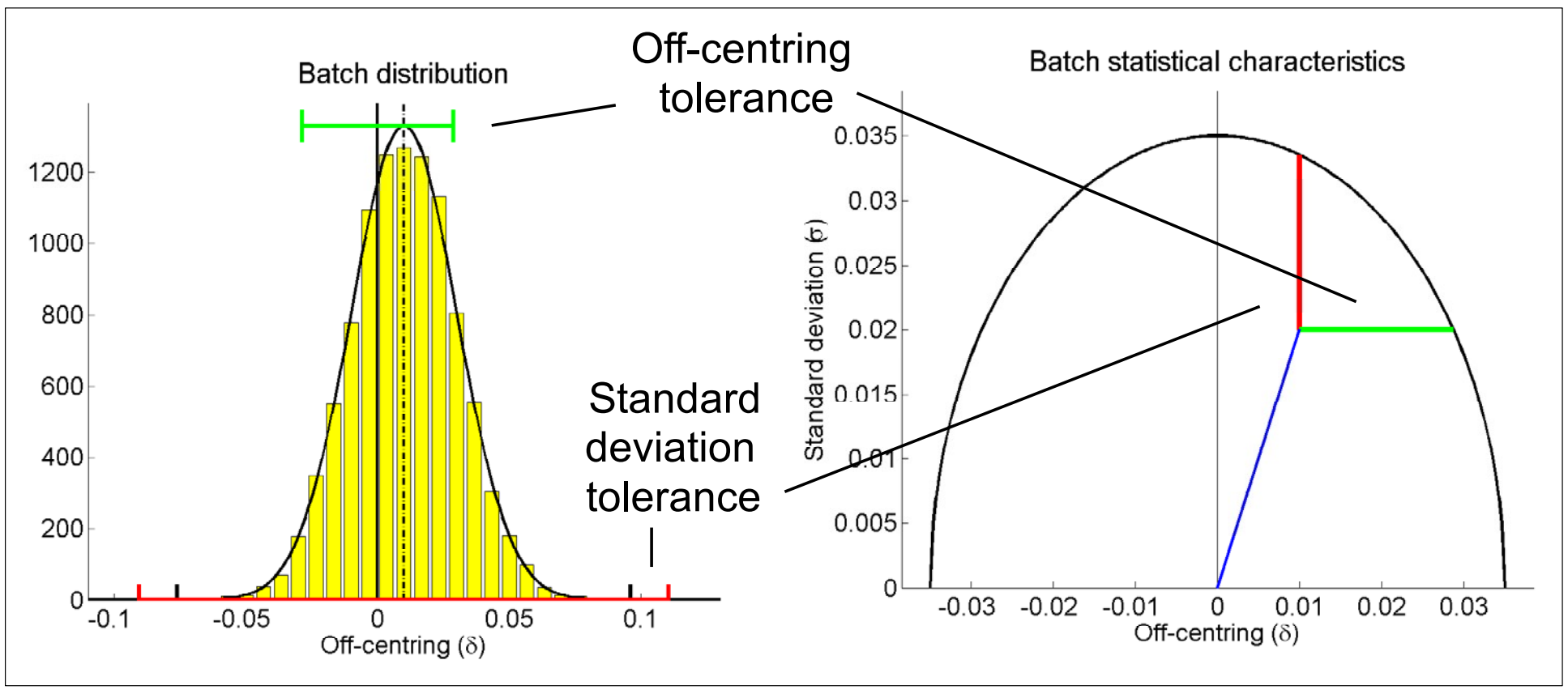

## 2.1.2 The inertial tolerancing method

Inertial tolerancing is a statistical method of tolerance synthesis. The allocation strategy is similar to that of the traditional statistical tolerancing. Let us consider components with centred batches. The resultant assembly is then a centred batch. The functional requirement is defined by a tolerance interval that is supposed to contain six standard deviations of the resultant batch. As the component inertias in the case of centring are given by the batch standard deviations, the inertial tolerances of components under the assumption of independent variables are given by:

$$I_i = \frac{R_0}{6.\sqrt{n}} \quad (5)$$

where:

$I_i$ = the inertial tolerance of the i-th components

$R_0$ = the tolerance interval of the functional requirement

n = the number of components in the dimensions chain.

In the general case, the inertial tolerance allocation is given by:

$$I_i = \frac{{}_i . R_0}{6.\sqrt{{}^{22}_{ii} .}} \quad (6)$$

where $_i$ corresponds to influential coefficients of the i-th component on the resultant assembly and $_i$ is the feasibility index of the i-th component that allows a nonuniform distribution of the tolerances.

## 2.1.3 Improvement of the approach

Although the allocation of tolerances is made by a statistical approach, the 1D inertial tolerancing does not offer the disadvantages of the traditional 1D statistical tolerancing discussed by Graves and Bisgaard (2000). Moreover, Parlar and Wesolowsky (1999) discusses the interests of using the Taguchi capability index Cpm for the tolerance interval, which is equivalent to the Cpi index for inertial tolerance. Some of our works (Adragna et al., 2006a; 2007a) show that the inertial tolerancing allows guaranteeing the functional requirement given by a tolerance interval $R_0$ and a Cpk index $Cpk_{FR}$, or a tolerance interval $R_0$ and a nonconformity rate $NCR_{FR}$. A very simple relation allows defining the components Cpi capability indices to guarantee the functional resultant $Cpk_{FR}$ index:

$$Cpi = \sqrt{Cpk_{FR}^2 + \frac{n}{9}} \quad (7)$$

where n is the number of components in the dimensional chain. The link between the component Cpi index and the maximum limit of the functional requirement $NCR_{FR}$ is given by the resultant distribution law, chosen as Gaussian, under the consideration of off-centred distribution.

### 2.1.4 Theoretical application

Let us consider a 1D tolerancing problem of a three-part stackup. The functional requirement is the total dimension of the assembly. The functional tolerance is chosen as t = 0.2 mm and the maximum nonconformity rate is $NCR_{FR}$ = 500 parts per million (ppm). Under the assumption of Gaussian distribution of the resultant assembly, the $Cpk_{FR}$ index corresponding to $NCR_{FR}$ is $Cpk_{FR}$ = 1.10. Thus the component Cpi index guaranteeing FR is Cpi = 1.24, given by Equation (7).

The histogram in Figure 2 shows the obtained NCR of the resultant assembly for components with only positive off-centring and acceptable Cpi capability indices. Although these assumptions are unrealistic and pessimistic, one can observe that the maximum NCR does not exceed the $NCR_{FR}$

Figure 2 Application and NCR results (see online version for colours)

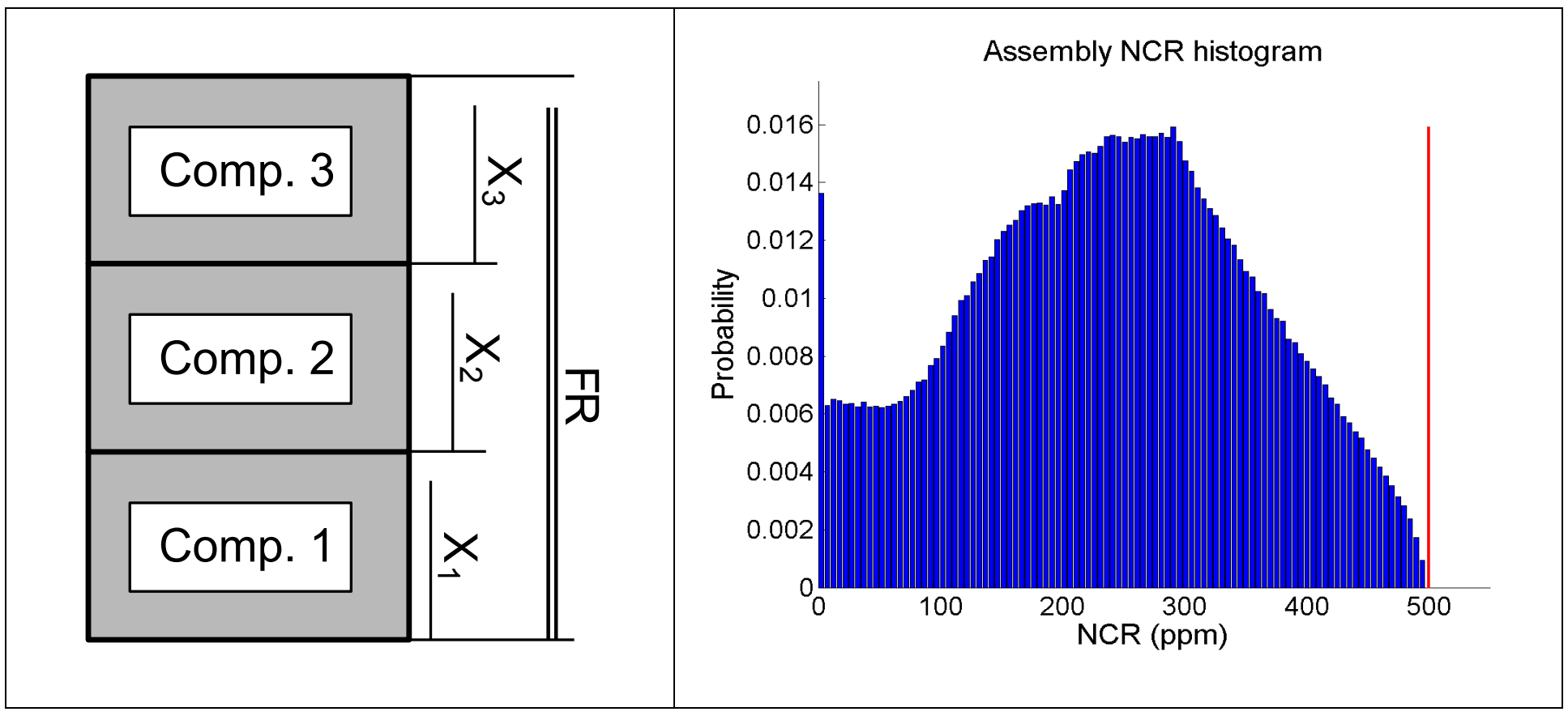

## 2.2 The modal characterisation of form deviations

This part presents a method to characterise the deviations of measured shapes. It is then possible to identify rigid deviations such as location and orientation, and several form deviations.

### 2.2.1 Origin of modal characterisation

The description of form deviations is a large domain for which one can find lots of propositions for specific applications, such as the famous Fourier transform applied to roughness filtering, the Zernicke polynomials used for the qualification of lens deviation, the Discrete Cosinus Transform (DCT) proposed by Huang and Ceglarek (2003), and a combination of the Chebyshev polynomials and Fourier series used by Summerhays et al. (2002) to characterise the form deviations of cylinders. Although these approaches are efficient for the creation of a deviation basis and the analysis of form, they are applied on some specific shapes (profile, discus, rectangle or cylinder).

Introduced by Samper and Formosa (2007), the modal characterisation of shape deviations is a generic approach that is able to describe form deviations of any geometry. Its main advantage is its form defects basis for the characterisation of measured form deviations. Each mode of the basis is defined by the shapes of the modal vibration of the ideal geometry to analyse. For practicality, the ideal geometry is discretised with nodes corresponding to measured nodes. Deformed shapes basis are natural mode shapes of the Finite Element Model (FEM) of the nominal geometry. This approach allows obtaining a modal basis for any shape, such as the gap flush profile of an upper hood (Adragna et al., 2006b) or its surface (Favrelière et al., 2007a), the external cylinder of a hydraulic spool (Adragna et al., 2006c) and a sphere (Favrelière et al., 2007b).

In our application, the surfaces are the contact surfaces $B_i$ and $C_i$ and the functional requirement surfaces $A_i$ which are rectangular planes. Although the modal vibration resolution of a rectangular plane can be analytically expressed, our solution is given by a FEM solver. As a good practice, some nodes' degrees of freedom are fixed:

- the translation degrees of freedom along the plane axes defining membrane vibration
- the rotation of nodes around the plane's normal direction.

This avoids useless modes (form deviations) for the qualification of form deviation expressed along the plane's normal direction. Thus, a natural modal basis is obtained composed of three rigid body modes, originally expressed by the combination of rigid translation and rotations, while the other modes are form deviation shapes.

### 2.2.2 Evolution of the modal characterisation

Metric modes

The first modification of the modal basis consists of specifying the modes' amplitude to the unit. This allows linking the modal coefficient to the exact amplitude of the mode. Detailed in a further part, the characterisation of a measured deviation shape gives a modal signature containing all modes' influences for this analysed form deviation. The modal coefficients correspond to the influence of each mode for the description of the shape, and their values are expressed in the unit of the measured form deviation (mm, µm or inch).

Rigid modes

The second modification of the natural modal basis consists of specifying the rigid modes. In our case, the first rigid mode, $_1$, is defined as a location deviation that corresponds to identical translation of the nodes along the plane's normal direction. The second and third rigid modes, $_2$ and $_3$, can be interpreted as rotation deviations around chosen axes that are symmetrical axes of the plane.

This parameterisation can be compared to and translated into the SDT parameterisation. Figure 3 shows two of the three rigid deviation modes of the modal basis and two different SDT data positioned on the plane. These models can be linked thanks to linear relations expressed as a matrix.

Figure 3 Some of the shapes of the modal basis (see online version for colours)

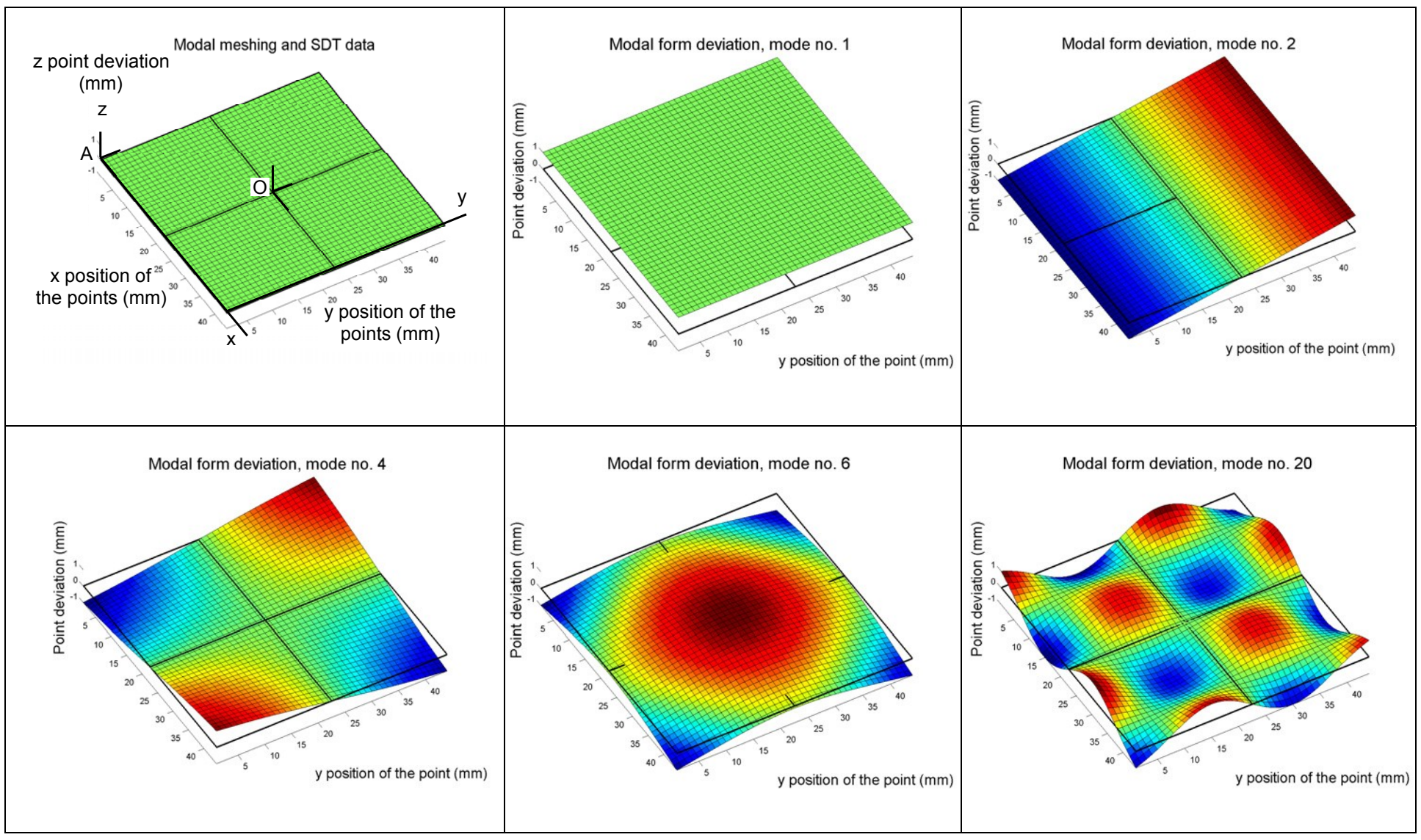

For the SDT datum centred on the O point that corresponds to the symmetric centre of the plane, the transition matrix from the rigid modal coefficients to the SDT components is diagonal:

$$\begin{bmatrix} T_z \\ R_x \\ R_y \end{bmatrix}_O = [SDT]_O \cdot \begin{bmatrix} 11 \\ 22 \\ 33 \end{bmatrix} = \begin{bmatrix} 1 & 0 & 0 \\ 0 & \frac{2}{L_y} & 0 \\ 0 & 0 & \frac{2}{L_x} \end{bmatrix}_O \qquad (8)$$

For the second SDT datum centred on the A point, the transition matrix is no longer diagonal:

$$\begin{Bmatrix} T_z \\ R_x \\ R_y \end{Bmatrix}_A = [T_{SDT}]_A \cdot \begin{Bmatrix} \lambda_1 \\ \lambda_2 \\ \lambda_3 \end{Bmatrix} = \begin{bmatrix} 1 & \frac{1}{L_y} & \frac{1}{L_x} \\ 0 & \frac{2}{L_y} & 0 \\ 0 & 0 & \frac{2}{L_x} \end{bmatrix}_A \quad (9)$$

This last matrix can be diagonal by changing the rigid modes' shapes.

### 2.2.3 Application of the method

Figure 3 shows some of the first modes of the basis. The first image presents the meshing of the plane and the location of two SDT data used in the previous part. Then, two of the three rigid modes are shown. They can be named translation and rotations around the x axis. The other modes correspond to form deviations. Mode no. 6 can be classified as domed shape, mode no. 4 can be called chip mode, and other modes such as mode no. 20 can be classified as undulation modes.

A measurement D is characterised in the modal basis B. Because the modal basis B is not an orthogonal and normal basis, the $\lambda_i$ coefficients are identified by the dual basis. The modal signature, vector of the modal coefficients $\lambda$, is then computed by:

$$\lambda = ((B^t B)^{-1} B^t) . D \quad (10)$$

As the deviation shape is characterised by the modal signature $\lambda = [\ldots \lambda_i \ldots]$, the form deviation can be recomposed and the efficiency of its characterisation can be evaluated. The recomposed, or reconstructed, shape R of the form deviation is given by:

$$R = B . \lambda \quad (11)$$

The reconstruction error $\varepsilon$ is given by the form difference between the measured shape D and its reconstructed shape R. One can then compute the scalar residue r given by the norm of the reconstruction error. It is even possible to define r(m), the scalar residue of the modal characterisation with m modes. One then defines a relative scalar $\rho(m,n)$ that is the ratio of the r(m) scalar residue over r(n). This finally allows comparing the efficiency of one mode or a lot of modes in the whole characterisation.

## 2.3 3D inertial tolerancing: fusion of the modal characterisation and the inertial criterion

### 2.3.1 The 3D inertial acceptance criterion

Presented in Adragna et al. (2007b), the inertial acceptance criterion is applied to the 3D and form characterisation method, which is the modal parameterisation. Pillet et al. (2005) introduced the inertial criterion on geometrical deviations. They hence define the inertia of a measurement set corresponding to deviation points. In the case of n surfaces measured by k deviation points, the surface batch inertia is defined as:

$$I = \sqrt{\frac{1}{n}\frac{1}{k}\sum_{j=1}^{n}\sum_{i=1}^{k}\left(X_{i,j} - \bar{X}\right)^2} \quad (12)$$

where $X_{i,j}$ is the measurement of the i-th point on the j-th shape. The second sum corresponds to the computation of the inertia of the i-th point defined by Equation (1). Hence, the original definition of the surface batch inertia corresponds to the quadratic mean of the surface point inertias:

$$I = \sqrt{\frac{1}{n}\sum_{j=1}^{n} I_j^2}. \quad (13)$$

### 2.3.2 Fusion into the modal parameterisation

As the modal parameterisation is a linear relation (Equation 11), Saporta (2000) gives two statistical relations:

$$\propto_R = B.\propto \quad (14)$$

where $\propto$ corresponds to the mean modal signature of a set of modal signatures $\lambda_i$ and $\propto_R$ corresponds to the mean shape. This mean shape indicates the mean deviation of each point of the characterised surface batch:

$$\Sigma_R = B.\Sigma.B^t \quad (15)$$

where $\Sigma_R$ is the covariance matrix of the surfaces points and $\Sigma$ is the covariance matrix of the modal coefficients. Then, the standard deviation form of the characterised surfaces is the square root of the $\Sigma_R$ diagonal. In relation to these statistical characterisations of a form deviation batch, Favrelière et al. (2007b) gives graphical representations of these mean and standard deviation shapes.

The inertial criterion of a form deviation batch described by the modal parameterisation can then be expressed as follows:

$$I = \sqrt{\frac{1}{n}\sum_{j=1}^{n}\left((B.\propto)^2 + diag(B.\Sigma.B^t)\right)} \quad (16)$$

where k is the number of measured points on the surfaces and is also equal to the length of the mean shape $B.\propto$. The diagonal of the recomposed shapes covariance matrix $B.\Sigma.B$ is the variance vector of length n.

The advantages of using the modal parameterisation to compute the surface batch inertia is the reduction of the parameter number from k points to m modal coefficients with m<k, and the knowledge of the form deviation (chip, domed, etc.).

### 2.3.3 Detection of a disadvantage and a solution

Highlighting of a problem

As the surface inertia is the quadratic mean of point inertias, some configurations may cause problems. Let us consider a 3D model based on the location and orientation deviation modes applied to the theoretical example presented in Figure 2. This problem can also be treated as four 1D tolerancing problems at each part's border. Then, the 3D FR is guaranteed if all 1D FR are guaranteed.

To underline the disadvantage of the initial geometrical inertia, let us consider a case where both left 1D dimension chains are perfect (centred and not dispersed) but the right dimension chains have variations. In this theoretical case, the surface inertia of the lower part I is defined by the quadratic mean of both front and rear left inertias, $I_{lf}$ and $I_{lr}$, and both front and rear right inertias, $I_{rf}$ and $I_{rr}$ considered as equally deviated:

$$I = \sqrt{\frac{1}{4}\left(0 + 0 + I_{rf}^2 + I_{rr}^2\right)} = \frac{I_r}{\sqrt{2}}. \quad (17)$$

Thus, the right corner variations are larger than the surface inertial tolerance. In this condition, the right dimensional chains do not guarantee its FR, thus the assembly FR is not guaranteed.

Solution: the 3D adjusted inertial criterion

To solve this problem, and in the assumption of no form deviation (only location and rotation deviations), the adjusted inertial criterion is proposed (Adragna et al., 2007b). The aim is to consider only the worst points inertia in order to avoid the compensation detailed in the previous part. The adjusted inertia is then defined as:

$$I_{adj} = \mathrm{Max}(I)_j \quad (18)$$

This new definition applied on the 3D problem with modal parameters can be written as:

$$I_{adj} = \sqrt{(|\ _1| + |\ _2| + |\ _3|)^2 + \ _1^2 + \ _2^2 + \ _3^2} \quad (19)$$

With the SDT datum centred on the O point, a similar definition can be found. The maximum deviation is given by:

$$|\ |_{Max} = |T_z| + |R_y|\frac{l_x}{2} + |R_x|\frac{l_z}{2} \quad (20)$$

The inertia is then given by:

$$I_{adj} = \sqrt{\left(|\ _{Tz}| + \frac{l_x}{2}|\ _{Ry}| + \frac{l_z}{2}|\ _{Rx}|\right)^2 + \ _{Tz}^2 + \left(\frac{l_x}{2}\right)^2 \ _{Ry}^2 + \left(\frac{l_y}{2}\right)^2 \ _{Rx}^2} \quad (21)$$

One can identify the following combination law where $_{Ry} = \frac{l_x}{2}$ and $_{Rx} = \frac{l_y}{2}$:

$$I_{adj} = \sqrt{\left(\delta_{Tz} + \delta_{Ry} \cdot y + \delta_{Rx} \cdot x\right)^2 + \sigma_{Tz}^2 + \sigma_{Ry}^2 \cdot y^2 + \sigma_{Rx}^2 \cdot x^2} \quad (22)$$

The definitions Equation (19) and Equation (21) are equivalent and the transition matrix of Equation (8). One can also note the assumption of an independent parameter due to the absence of the covariance. The covariance is not considered for now in these two relations, but is implicitly taken into account in the relation (18).

In order to use the relations Equation (19) and Equation (21) in a valid context, let us consider the assumption of independent parameters. In this case of 3D parameterisation, it is possible to represent the geometrical inertial deviation. We chose to represent the inertial deviation and not the inertial tolerance as for the 1D representation in Figure 1. Two kinds of inertial deviations are identified:

1 the mean deviations defining a triangular domain

2 the standard deviations defining an elliptic domain.

Figure 4 presents a batch of form deviation, its representation in the domain of the first three modes and its inertial deviation in the same modal domain.

Figure 4 (a) Batch of form deviations, (b) modal representation of the batch, (c) inertial deviation domains of the modal components (see online version for colours)

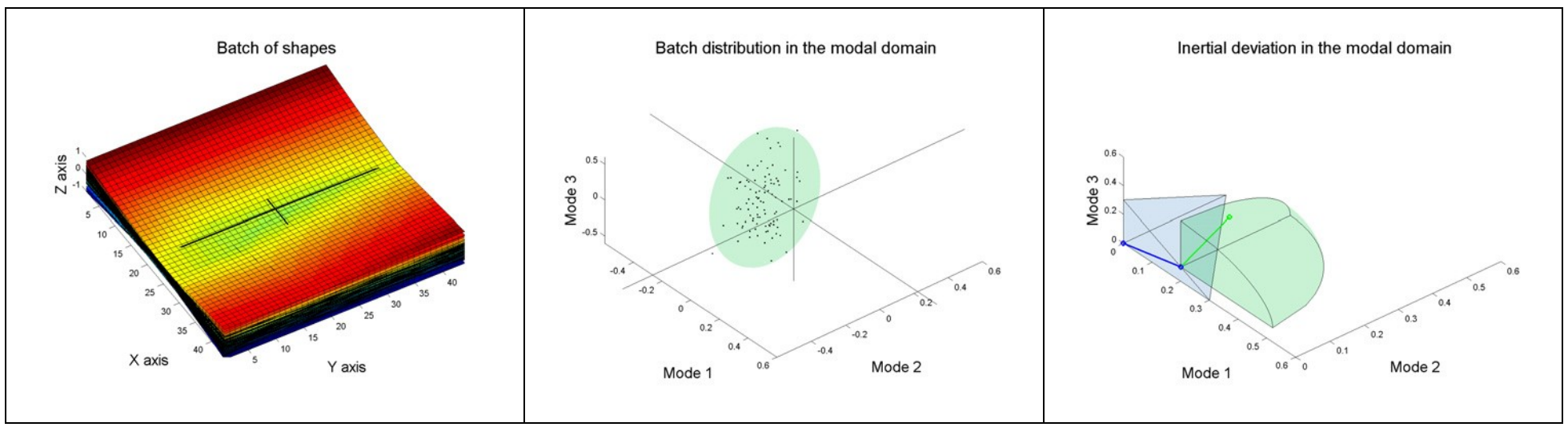

One can note from the first image of the batch shapes that the rotation around the y-axis (mode 3) is the largest dispersion. This can be observed on the second representation of the batch in the modal domain. The largest direction of the ellipsoid is along the mode 3 axis. Finally, the last representation of the batch deviation shows that the maximum mean deviation is due to the translation (mode 1), completed with a little rotation (mode 3) that is observable in the blue tetrahedron. The standard deviation presented in the part of the green ellipsoid shows an important standard deviation on mode 3, and then on mode 2.

## 3 3D inertial tolerancing without form errors

This part proposes a statistical tolerancing approach based on the 3D adjusted inertia and the consideration of only rigid shapes (no form deviation). The case of application is the three-part stack-up with a lever arm, and FR is a location tolerance zone t = 0.2 mm. The statistical approach is compared to the worst-case tolerancing on a common model: the SDT, the data origins of which are placed on each face's centre.

Figure 5 (a) Batch of form deviations, (b) modal representation, (c) inertial deviation domains (see online version for colours)

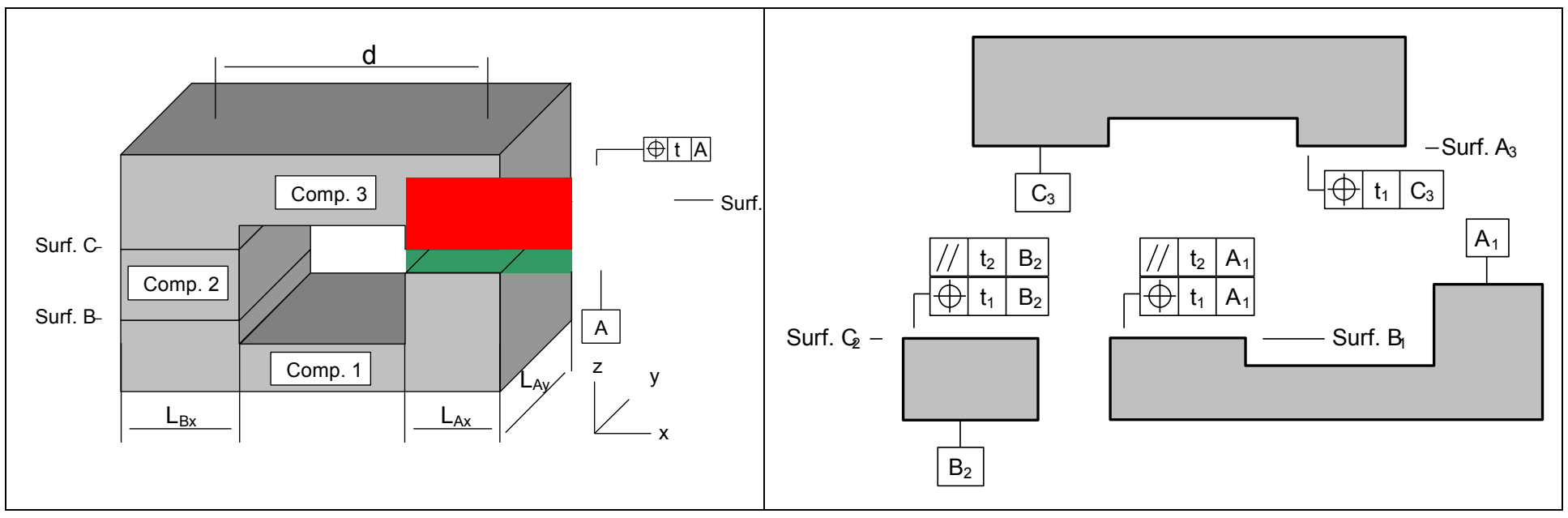

The qualitative tolerancing is arbitrarily chosen as follows:

- a location tolerance zone applied on the three surfaces A, B and C. The size of the tolerance is $t_1$
- an orientation tolerance zone applied on the surfaces B and C. The size of this tolerance is $t_2$
- $t_1$ is twice as large as $t_2$.

The mechanism characteristics are $L_{Bx}$ = 80 mm, $L_{Ax}$ = 100 mm, $L_{Az}$ = 80 mm and d = 220 mm.

The quantitative tolerancing with a worst-case approach gives $t_1$ = 0.0229 mm and $t_2$ = 0.0114 mm. This result is found with the deviation domain method (Giordano et al., 2001).

## 3.1 A proposition of 3D inertial tolerancing

Our proposition of inertial tolerancing consists of identifying dimensional chains for each component of the SDT. In the treated case, three dimensional chains are identified:

1 translation of each surface on the y axis

2 rotation of each surface around the x axis

3 rotation of each surface around the z axis.

These three dimensional chains give the maximum inertia on each axis of the SDT for each component. In order to have the same restricted tolerance distribution as the worst-case tolerance, it is chosen to not uniformly distribute the tolerance on the rotations' components. As with the worst case, the rotation tolerances of component 3 are twice as large as the rotation tolerances of components 1 and 2.

### 3.1.1 Allocation of the inertial tolerances

Inertial tolerancing of translations

The dimension chain of the translations on the z axis is:

$$T_{Max} = T_{z1} + T_{z2} + T_{z3}. \quad (23)$$

Thus the inertial tolerancing relation with the assumption of uniform distribution of the tolerance is:

$$I_{Tzi} = \frac{t}{6.\sqrt{3}}. \quad (24)$$

The components inertial tolerances are $I_{Tzi} = 0.0192$ mm.

Inertial tolerancing of rotation around x

The dimension chain of the rotations around the x axis is:

$$_{Max} = R_{x1}.\frac{L_{Ay}}{2} + R_{x3}.\frac{L_{Ay}}{2} + R_{x3}.\frac{L_{Ay}}{2} \quad (25)$$

The tolerance distribution uses the feasibility indices $_1 = _2 = 1$ and $_3 = 2$, thus:

$$I_{Rxi} = \frac{_i.t}{3.L_{Ay}.\sqrt{\sum_i {_i}^2}}. \quad (26)$$

The components' inertial tolerances are then $I_{Rx1} = I_{Rx2} = 3.4.10^{-4}$ rad and $I_{Rx3} = 6.8.10^{-4}$ rad.

Inertial tolerancing of rotation around z

The dimension chain of the rotations around the y axis is:

$$_{Max} = R_{y1}.d + R_{y2}.d + R_{y3}.\frac{L_{Bx}}{2} \; _y \frac{L_{Bx}}{2} \; _y \frac{}{Ax} \quad (27)$$

The tolerance distribution uses the feasibility indices $_1 = _2 = 1$ and $_3 = 2$, thus:

$$I_{Ryi} = \frac{_i.t}{6.\sqrt{2.{_{1}}^2\left(\frac{L_{Bx}}{2}+d\right)^2 + {_{3}}^2 \left(\frac{L_{Bx}}{2}\right)^2 . \frac{}{Ax}}}. \quad (28)$$

The components' inertial tolerances are then $I_{Ry1} = I_{Ry2} = 8.4.10^{-5}$ rad and $I_{Ry3} = 1.7.10^{-4}$ rad.

Identification of combination laws

As the SDT components do not have the same tolerance, a combination law has to be determined. The inertial tolerance of the translation components is the larger tolerance; it is thus chosen as the reference tolerance. The influence of the two rotation components is identified by the ratio of the inertial tolerances. It is also possible to analytically identify these combination coefficients. The combination ratio is numerically identified as follows:

$_{Rx} \approx 56.6$ and $_{Rx} \approx 227.9$ for components 1 and 2

$_{Ry} \approx 28.3$ and $_{Ry} \approx 113.9$ for component 3.

### 3.1.2 Comparison to the worst-case tolerances

This part presents a comparison of the dispersion of both tolerancing approaches in the case of centred distribution of the components' deviations. Germain and Giordano (2007) shows that, in the case of a centred multinormal distribution of the components' deviations with six standard deviations within the tolerance interval on each SDT component, the filtering of parts out of the components' tolerance interval is not necessary due to the low resulting NCR on the resultant assembly. Since the comparison of a worst-case tolerance and a statistical tolerance would not be exact if different hypotheses are used, the same hypothesis is considered to compare the two tolerancing approaches.

Let us consider a multinormal distribution within the tolerance interval (six standard deviations within the tolerance zone for each SDT component). The parts distribution associated with the inertial tolerance has a similar repartition to the worst-case distribution, hence the comparison identifies the ratio of a homothetic transformation.

Figure 6 compares the multinormal distribution of components 1 and 2. The distribution is represented by a 3D ellipsoid representing +/–3 standard deviations. This representation allows an easier comparison; a set of points would have hidden the border of the worst-case tolerance domain. One can observe in this homothetic comparison that the inertial distribution is 63% larger than the worst case distribution.

Figure 6 Comparison of the tolerances of components 1 and 2, inertial distribution is 63% larger than worst case distribution (see online version for colours)

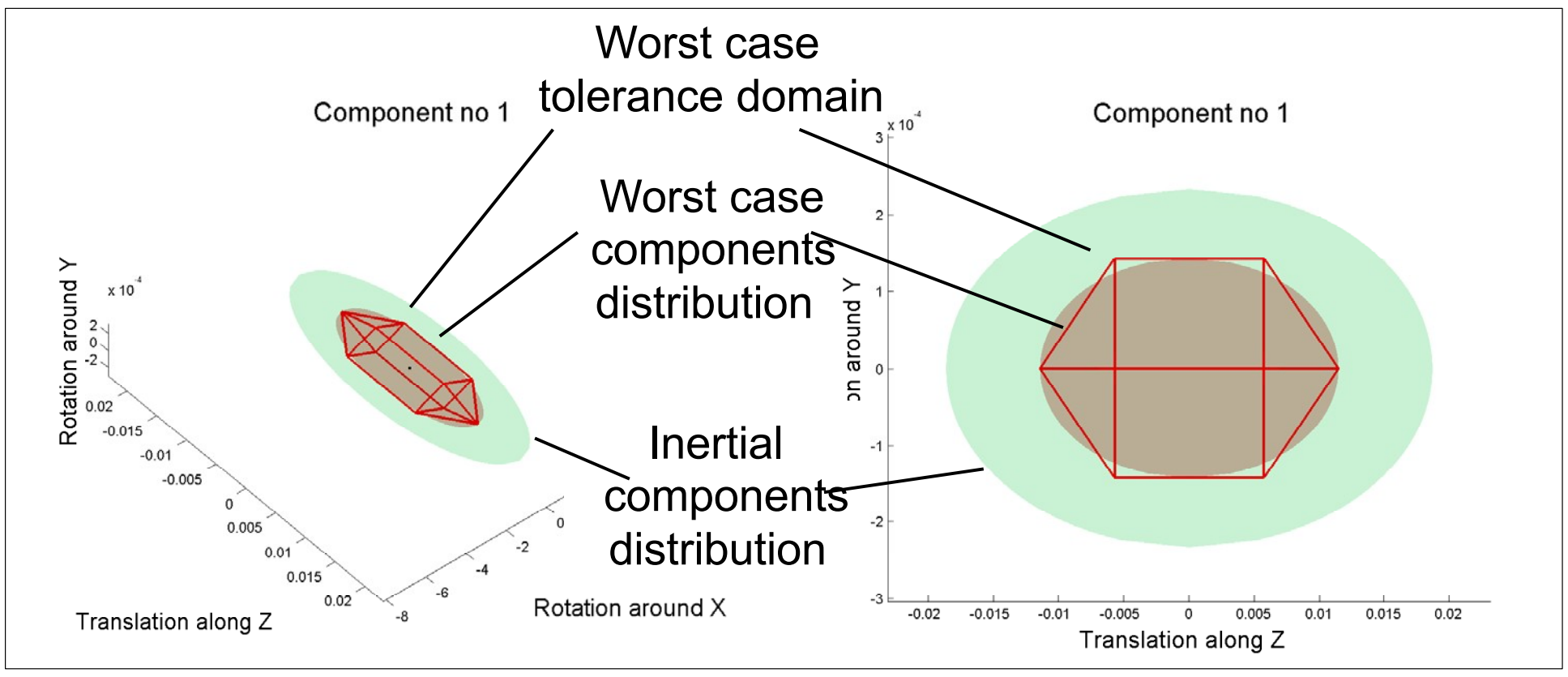

Figure 7 presents the comparison of the +/–3 standard deviation ellipsoids for component 3. In this case, the homothetic increase is 95% of the worst-case centred distribution.

The aim of tolerancing is to guarantee the final resultant assembly. In order to verify if the proposed tolerancing guarantees the respect of the FR, the characteristics (mean and covariance matrix) of resultant assembly are calculated as presented by Germain and Giordano (2007). Then 1 000 000 resultant assemblies are randomly drawn with a multinormal distribution and the resultant assembly covariance matrix. Figure 8 presents both the resultant assembly for the statistical assemblies based on the worst-case or inertial tolerances of the two previous figures. The dark points represent the assemblies' characteristics. The assemblies out of the FR (domain delimited by the blue lines) are included in the red circles.

Figure 7 Comparison of the tolerances of component 3, inertial distribution is 95% larger than worst case distribution (see online version for colours)

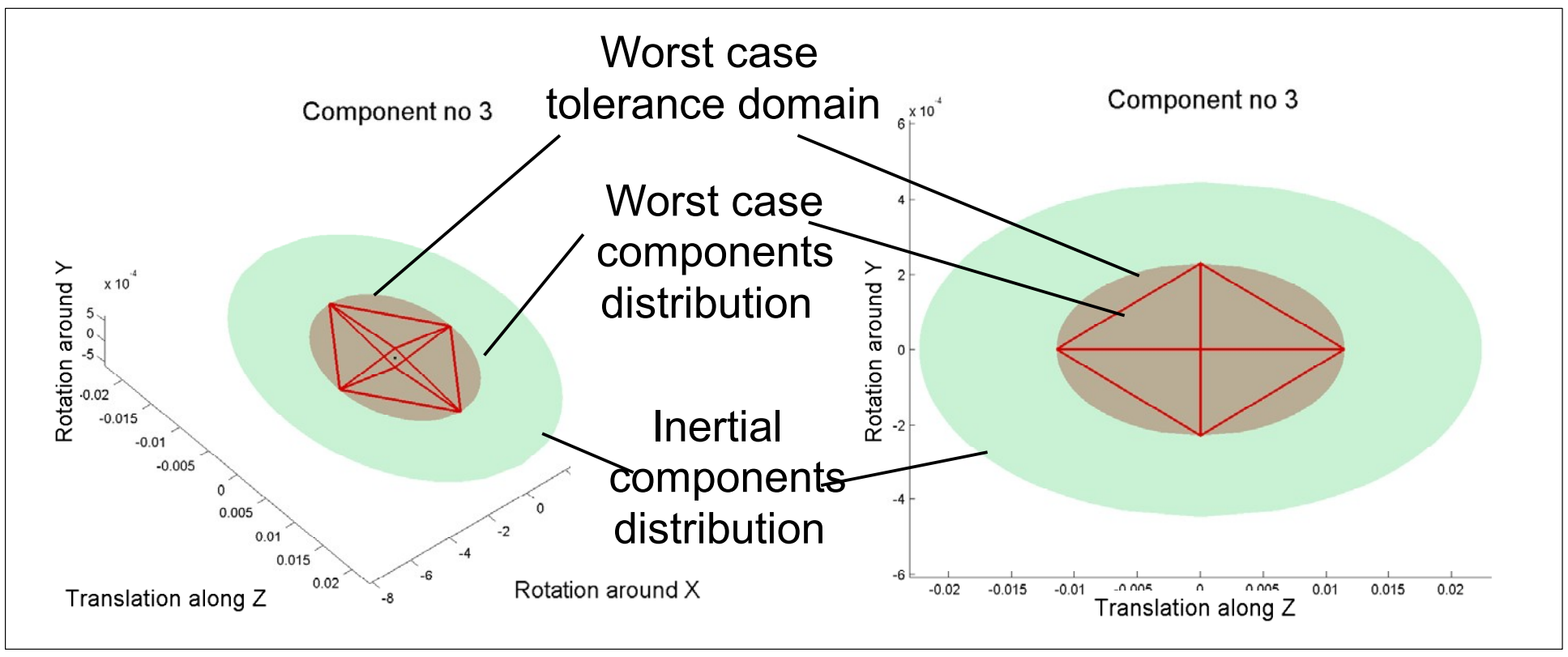

Figure 8 Resultant assembly distribution for component distributions centred in their worst-case or inertial tolerances (see online version for colours)

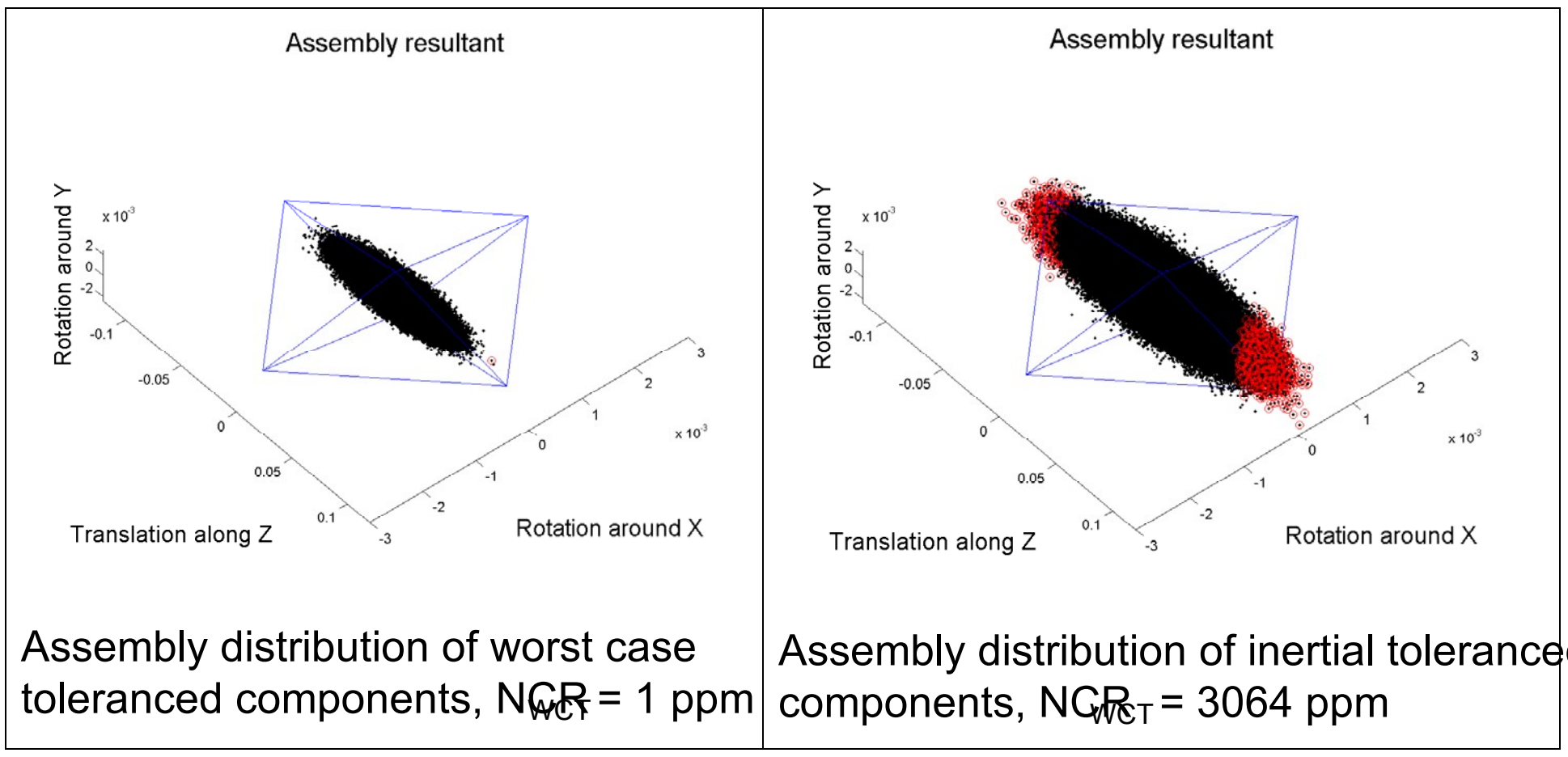

The NCR obtained for the statistical distributions based on the worst-case tolerancing is negligible; $NCR_{WCT}$ is a few ppm as presented by Germain and Giordano (2007). For the tolerances based on the inertial tolerancing, the $NCR_{In.}$ is around 3000 ppm = 3%. Although this resultant $NCR_{In.}$ is greater, it can be reduced by the use of the Cpi capability index. A summary of some assembly $NCR_{In.}$ in the case of centred components is presented in Table 1.

Table 1 NCR results of the resultant assembly for three different simulation cases

| Configuration | Cpi = 1 | Cpi = 1.16 | Cpi = 1.33 | Cpi = 1.5 |
|---|---|---|---|---|
| Centred | 1 (mean NCR) 3060 ( = 177) | 590 ( = 76) | 80 ( = 29) | 9 ( = 9) |
| Centred | 1 (worst NCR) 13 800 | 4700 | 1210 | 390 |
| Off-centred | 1 (worst NCR) 24 140 | 7270 | 1990 | 640 |

## 3.2 Analysis of the 3D inertial tolerances

The first simulation, presented as the configuration 'Centred 1' in Table 1, is illustrated in Figure 8 and corresponds to centred components with the standard deviation distribution detailed in Part 3.1.2. The second simulation, 'Centred 2' in Table 1, is illustrated in Figure 9 and considers centred components with a random standard deviation distribution.

The third simulation, 'Off-centred' in Table 1, is illustrated in Figure 10 and considers randomly off-centred components with a random distribution of the standard deviation. In order to avoid compensations of the components' off-centring and to evaluate a kind of worst configuration, components' off-centring are all considered positive.

Table 1 gives the results for the assembly NCR in the case of centred components and the maximum reached NCR with the hypothesis of random distribution of the components' standard deviation or random distributions of the component's off-centring and standard deviation.

The first line, Centred 1, corresponds to the particular case where all components are centred and their dispersion is similar to the one associated with the worst-case tolerances, presented in Figures 6–8. The indicated results are expressed in ppm. The NCR is the mean result of 2000 simulations and its standard deviation, also in ppm, is indicated. One can note that the observed standard deviation of the results follows the prediction of Cvetko et al. (1998):

$$\sigma_{NCR} = \sqrt{\frac{NCR(10^6 - NCR)}{N - 1}} \quad (29)$$

where N is the number of drawings.

For the two other lines, only one value is indicated that corresponds to the maximum NCR value obtained during our simulations. As the worst NCR is not easily obtained, components' off-centring are limited to a third of their inertial tolerance. Without this condition, the worst NCR is not drawn. To illustrate our purpose, Figure 11 shows two simulations for the Cpi capability index but different hypotheses on the components' off-centring. The histogram on the left is the result of random drawing of components' off-centring within a third of their inertial tolerance, and the second histogram is a random drawing of the off-centring within their entire inertial tolerance. One can note that the maximum NCR obtained in the second histogram is lower than the first one.

The second line, Centred 2, corresponds to centred components with random dispersion as presented in Figure 9. Although the multinormal distribution has a random allocation of its components' standard deviation, the capability indices of each components' batch is set to the specified value.

Finally, the last line, Off-centred, corresponds to the assembly simulation of randomly off-centred components and randomly distributed standard deviations. The assembly considers only positive off-centring of each component as presented in Figure 10.

From this table, one can conclude that the Cpi capability index has a real effect on the maximum obtained NCR. The proposed 3D inertial criterion seems to be a good criterion for 3D statistical tolerancing.

Figure 9 An assembly of centred components' batches, Cpi = 1.16, NCR = 633 ppm (see online version for colours)

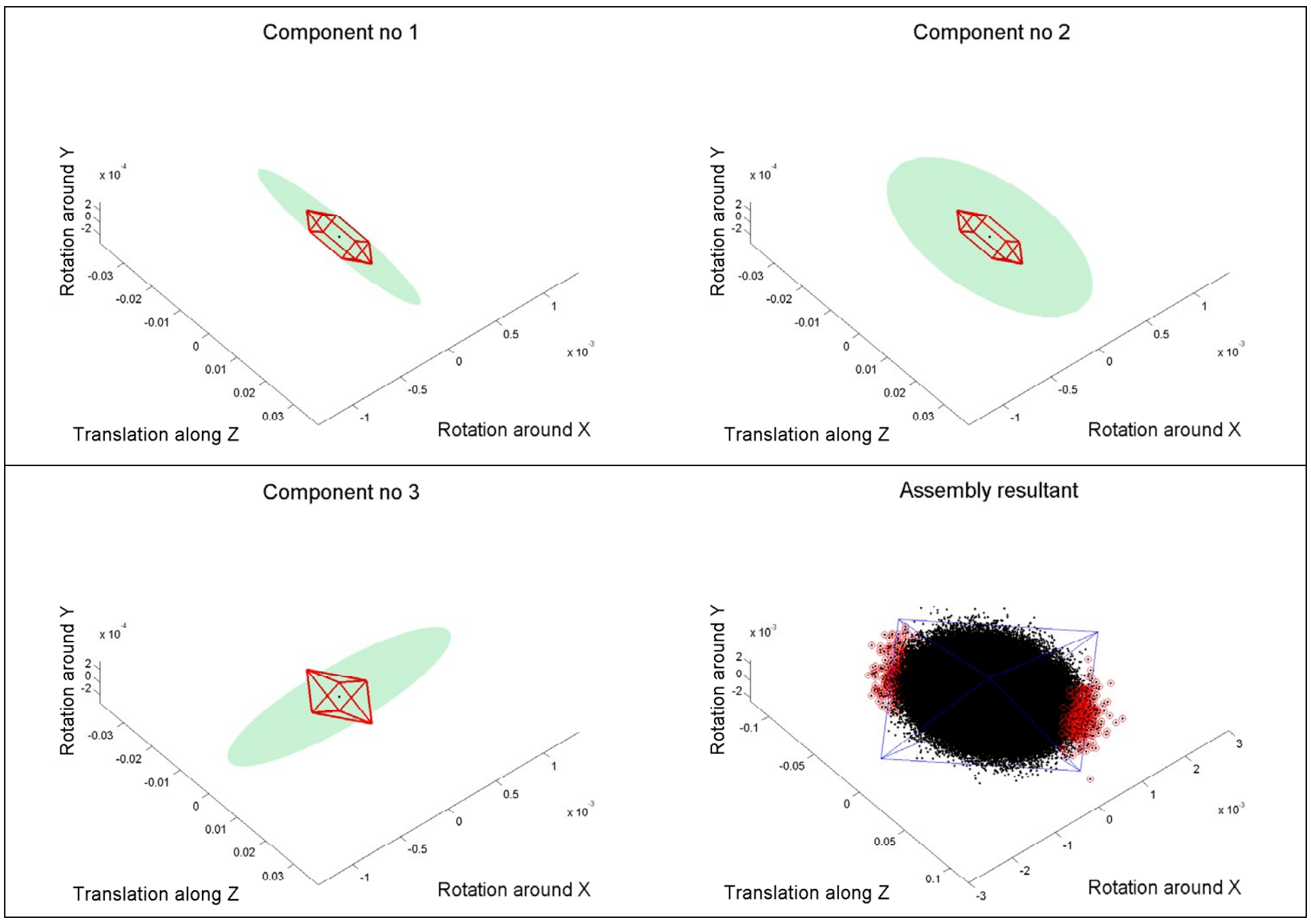

Figure 10 An assembly of randomly off-centred components' batches, Cpi = 1.16, NCR = 25 ppm (see online version for colours)

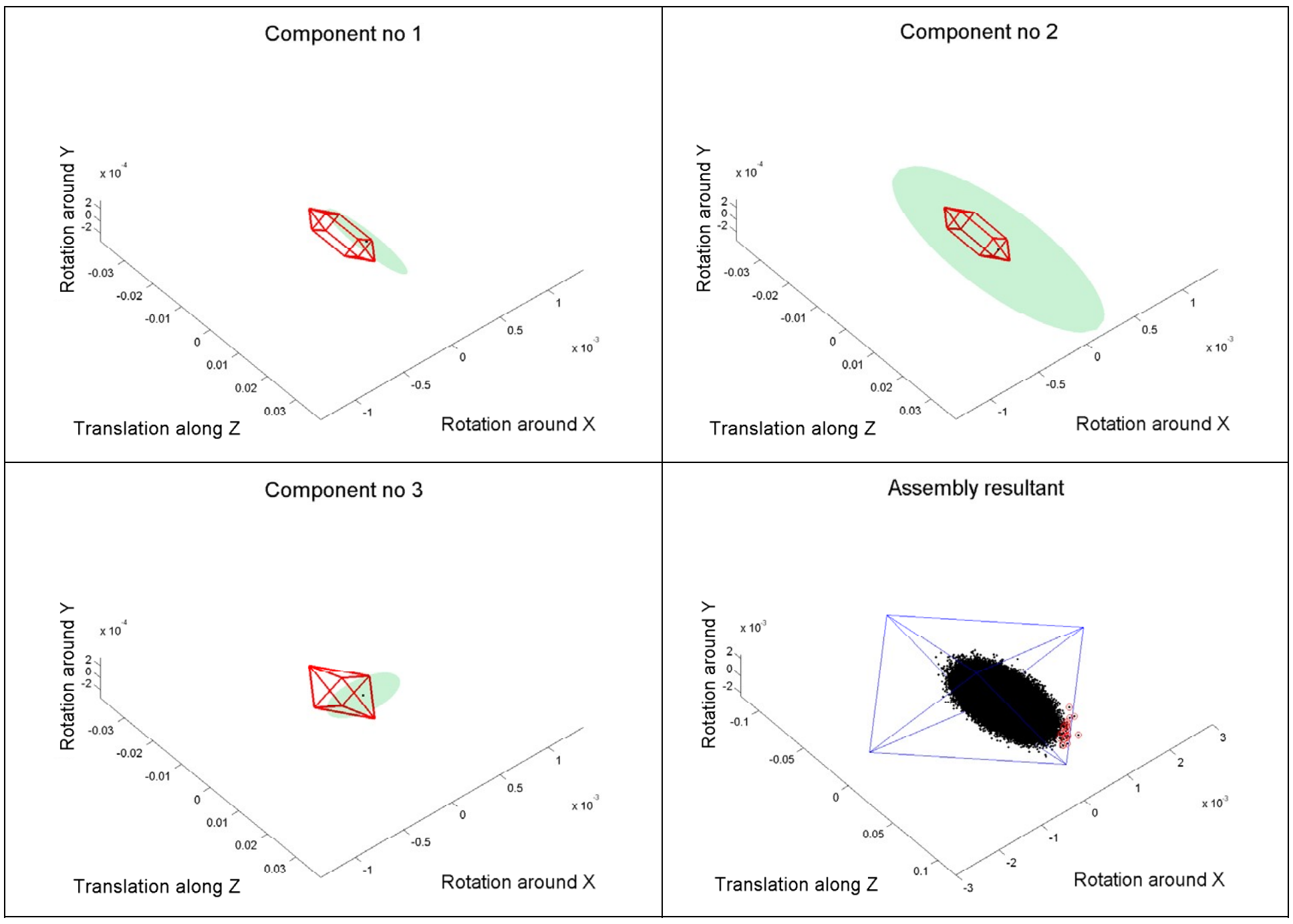

Figure 9 An assembly of centred components' batches, Cpi = 1.16, NCR = 633 ppm (see online version for colours)

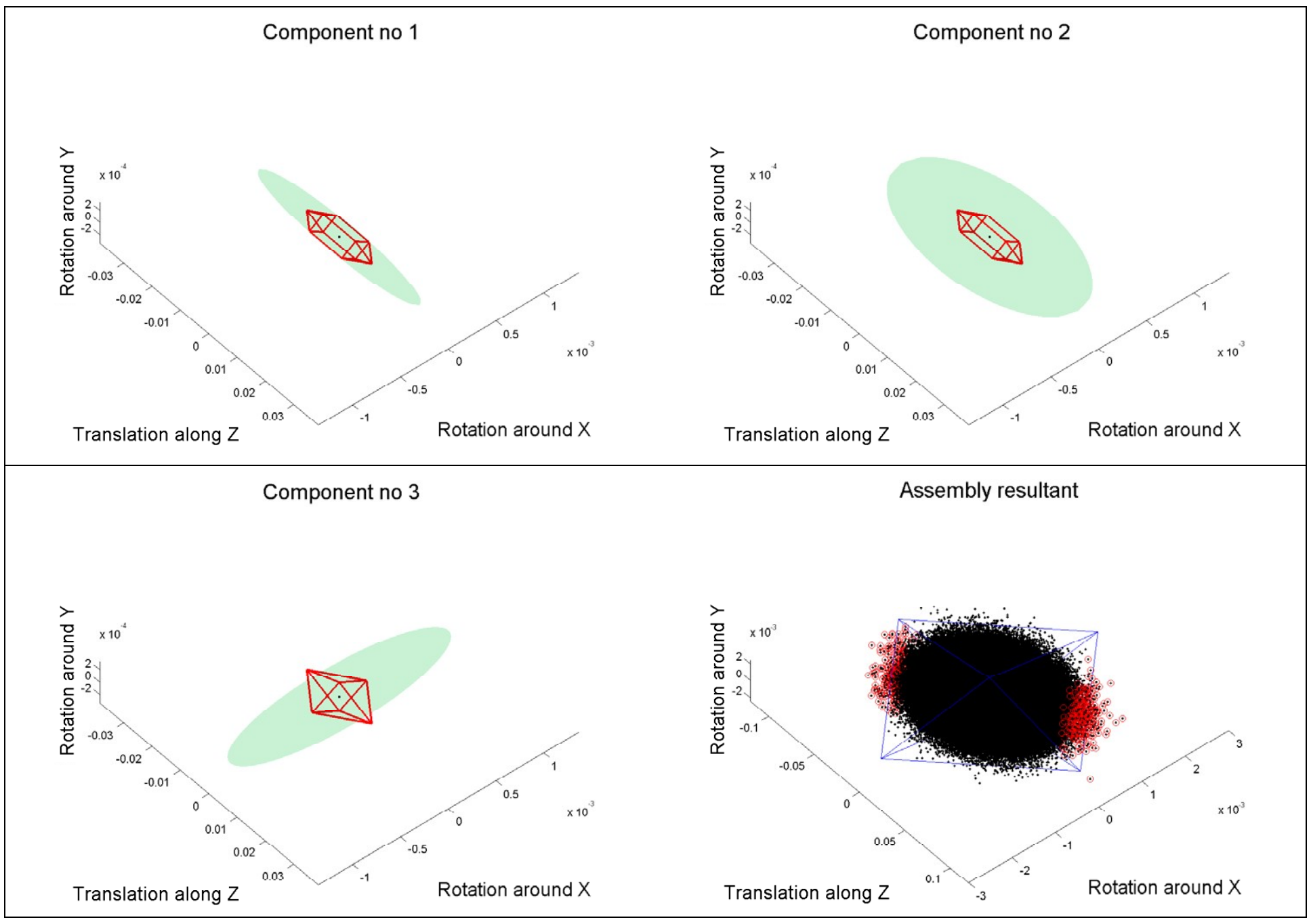

Figure 10 An assembly of randomly off-centred components' batches, Cpi = 1.16, NCR = 25 ppm (see online version for colours)

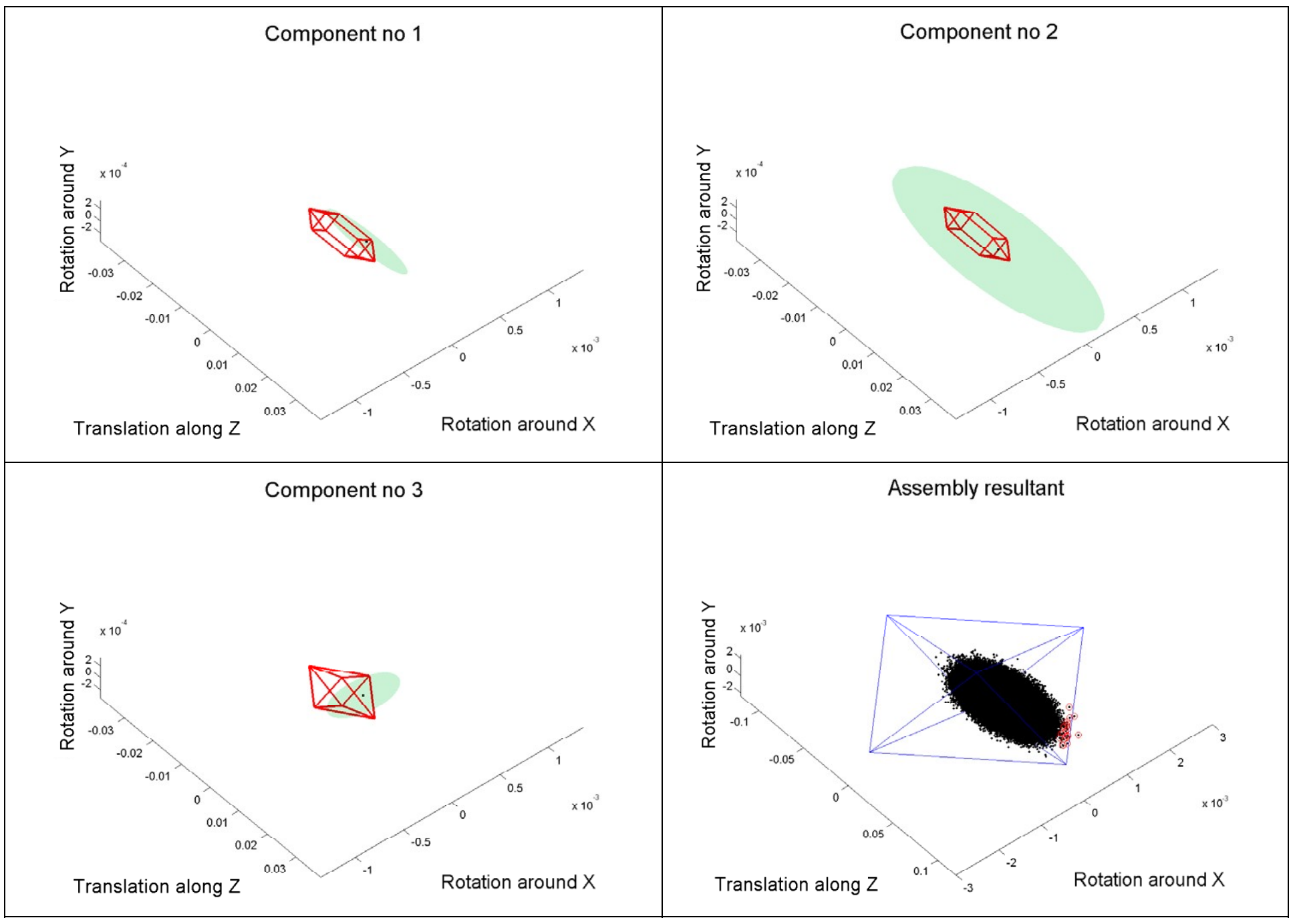

Figure 11 (a) random off-centring of components' in a third of their inertial tolerance, (b) random off-centring in their inertial tolerance (see online version for colours)

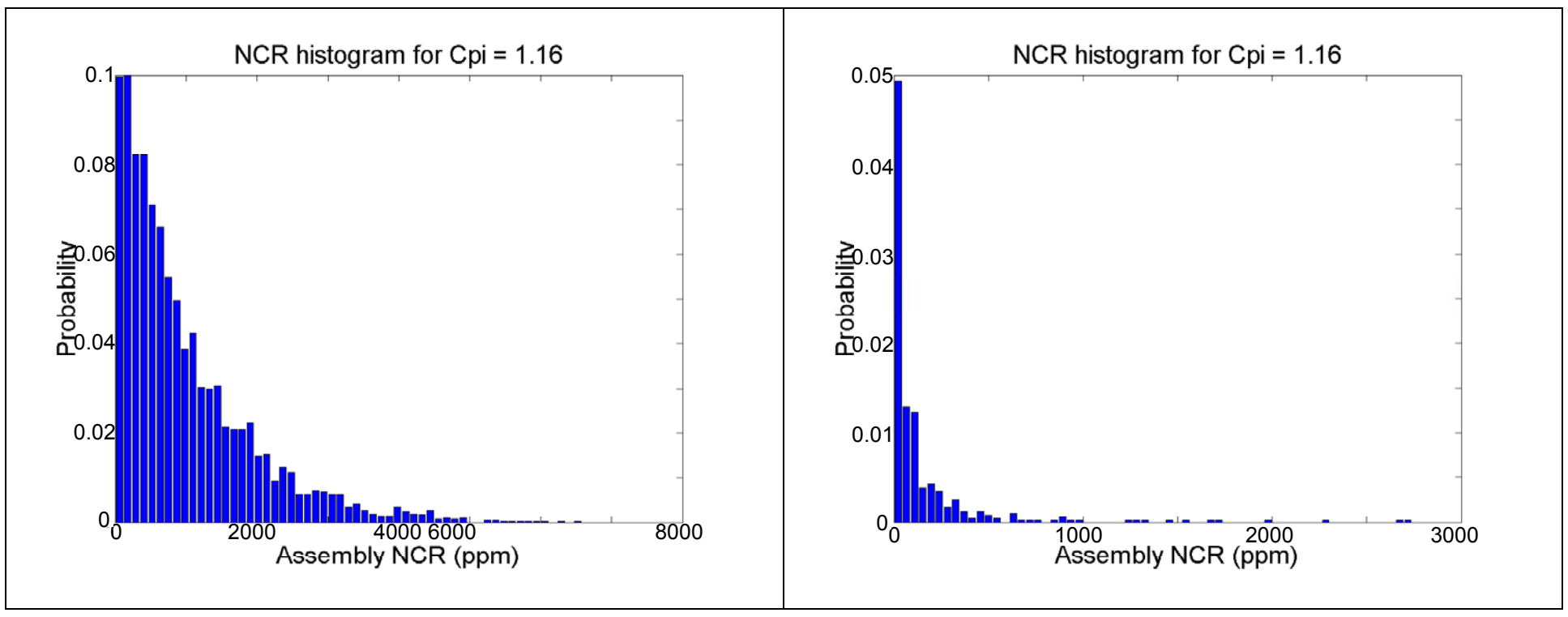

From this table, one can conclude that the Cpi capability index has a real effect on the maximum obtained NCR. The proposed 3D inertial criterion seems to be a good criterion for 3D statistical tolerancing.

## 4 Conclusion

This paper presents a new acceptance criterion for 3D statistical tolerancing. Derived from 1D statistical tolerancing, where it is efficient, the 3D inertia seems to be well adapted for the acceptance of rigid form deviations. The Cpi capability index has an influence on the resulting NCR. The results of this paper show why there is interest in this new acceptance criterion.

Some of our next works will aim to evaluate the components' configuration that give the worst assembly NCR. Then, as for the 1D statistical tolerancing, it may be possible to give the Cpi index that exactly guarantees the FR. Another objective is the consideration of form deviations and their specification in order to guarantee the FR.